# Automatic Removal of Cosmic Ray Signatures on Deep Impact CCDs

S. I. Ipatov[a,c], M. F. A'Hearn[a], and K. P. Klaasen[b]

[a]*University of Maryland, College Park, 20742-2421 MD, USA*
[b]*Jet Propulsion Laboratory/NASA, USA*
[c]*Present address: Department of Terrestrial Magnetism, Carnegie Institution of Washington, 5241Broad Branch Road, Washington, D.C. 20015-1305, USA; siipatov@hotmail.com*

**Abstract**

The results of recognition of cosmic ray (CR) signatures on a single image were analyzed for several codes written by several authors. For automatic removal of CR signatures on many images made during the Deep Impact mission, we suggest to use the code *imgclean* written by E. Deutsch, but other codes can be better for analysis of concrete images. *Imgclean* detects false CR signatures near the edge of a comet, and it often does not recognize all pixels of long CR signatures, but other codes considered sometimes does not come to the end. Our code *rmcr* is the only code among considered which allows to work with raw images. For most visual images made during low solar activity at exposure time $t>4$ s, the number of clusters of bright pixels on an image per second per sq. cm of CCD was about 2-4, both for dark and normal sky images. At high solar activity, it sometimes exceeded 10. The ratio of the number of CR signatures consisting of $n$ pixels obtained at high solar activity to that at low solar activity was greater for greater $n$. The number of clusters detected as CR signatures on a single infrared image is by at least a factor of several greater than the actual number of CR signatures; the number of clusters based on analysis of two successive dark frames is in agreement with an expected number of CR signatures. Some glitches of false CR signatures include bright pixels presented on different infrared images. Our interactive code *imr* allows a user to choose the regions on a considered image where glitches detected by *imgclean* as CR signatures are ignored. In other regions chosen by the user, the brightness of some pixels is replaced by the local median brightness if the brightness of these pixels is greater by some factor than the median brightness. The interactive code allows one to delete long CR signatures and prevents removal of false CR signatures near the edge of a comet.

## 1. Introduction

On July 4, 2005 the impactor (370 kg) of the Deep Impact (DI) spacecraft collided with Comet 9P/Tempel 1 at velocity of 10.2 km/s (A'Hearn et al., 2005). A lot of images were made during the flight of this spacecraft. The images were made by different cameras (Hampton et al., 2005). The high-resolution instrument (HRI) consists of an f/35 telescope with 10.5 m focal length, and a combined filtered CCD camera and infrared (IR) spectrometer. The medium resolution instrument (MRI) consists of an f/17.5 telescope with a 2.1 m focal length feeding a filtered CCD camera. The HRI and MRI are mounted on the flyby spacecraft. The third instrument called impactor targeting sensor (ITS) is a simple unfiltered CCD camera with the same telescope as MRI, mounted within the impactor spacecraft. All three instruments use a CCD with 1,024*1,024 active pixels.

A cosmic ray (CR) passing through the visual or IR detector can cause generation of a cluster of signal charge. For analysis of images, it is often needed to remove signatures of cosmic rays on the images. In the present paper, we consider the removal of CR signatures on DI images. This problem was discussed in a wide calibration paper by Klaasen et al. (2006). Below we pay more attention to the description of the codes used for removal of CR signatures and to their applications for the removal of CR signatures (especially on ITS and IR images) and pay less attention to statistics for CR signatures than in the above paper. The results were presented at several conferences (Asteroids, Comets, Meteors 2005; AAS 207 Meeting, COSPAR-2006, and 26th IAU General Assembly).

## 2. Codes for removal of cosmic ray signatures

The most reliable way to recognize CR events on astronomical images is to compare different images of the same region of the sky, but it is not always possible to do this. The previous codes for removal of CR signatures in a single image (e.g., *imgclean and crfind* written by E. Deutsch and R. White, respectively) were worked out for studies of typical sky images. Sometimes the above codes do not work well with visual images of a comet and with IR and raw (before calibration) visual images. The code *di_crrej* written by D. Lindler in 2005 for analysis of DI images has similar problems. These codes sometimes detect false CR signatures near the edge of a comet nucleus or on its coma and may have problems with long (oblique entry) CR signatures. *Crfind* and *di_crrej* only identify pixels corresponding to CR signatures, but do not replace these pixels. *Imgclean* uses the code *starchck* in order to make a decision whether a cluster corresponds to a star or to a cosmic ray. The latter code compares the brightness of the brightest pixel of a cluster with the brightness of several squares of close background and studies the location of the pixel in the cluster.

We wrote the code *rmcr* for removal of CR signatures. In contrast to the above codes, it can work also with raw (not only with calibrated) images, but the present version is slow if an image of a comet consists of a large number of pixels. The code *rmcr* analyses those pixels for which raw digital numbers (DNs) or calibrated radiances are greater than some limit *lim*. Only such pixels are considered to be caused by CRs. This limit can be an input parameter (e.g., for raw images) or it can be calculated as *lim=limit0*klim* (e.g., *klim*=3), where *limit0* is the median value of all pixels on an image. For calibrated images, one may not know *lim* in advance, so it is better to use the latter calculation of *lim*.

The code finds groups of pixels that are located close to each other. Pixels are considered to belong to one group if $(dx+1)^2+(dy+1)^2<ddilim$, where *dx* and *dy* are differences in coordinates *x* and *y* of two pixels (the width of a pixel is equal to 1). The default value of *ddilim* is equal to 8.1, i.e., two pixels separated by not more than one pixel belong to the same group. The code makes statistics for sizes of all groups. Sometimes *rmcr* was applied to pixels considered by other codes as belonging to CR events in order to get the distribution of the events over the number of pixels in one event.

*Rmcr* removes all clusters with $k_p<0.17$ (see the end of section 3). These clusters correspond to 'long' CR signatures. Charge clusters consisting of not more than *nlimit* pixels are also considered as CR signatures. Depending on a considered image and problem, the input parameter *nlimit* can be chosen to take any value (e.g., 10). In one version of the code, the clusters (exclusive for 'long' clusters) that are closer than *dss* (e.g., *dss*=10 pixels) to a defined rectangle that includes the comet and its coma are not considered as CR signatures. For small values of *nlimit*, it may be useful to run *rmcr* for calibrated images two times - first with a greater *lim*, and then with a smaller *lim*. The code runs slowly when there are a lot of pixels in all clusters (e.g., a comet occupies a considerable part of an image) because the code analyzes the entire image at once, rather than by small portions of the image at a time as do the other codes.

The *rmcr* code replaces detected CR signatures with values of their neighboring pixels. At the beginning of the IDL code we calculate *sky*=median(*ima, medima*), where *ima* is the initial image and the default value of *medima* equals 7. The brightness of the pixels belonging to CR events is replaced by *sky* values (if the *sky* brightness is less than the initial brightness).

The replacement of brightness of pixels belonging to signatures of CRs is incorrect in *imgclean*. Therefore the edge of an image of a comet after *imgclean* can be become 'torn' (though it must be smooth). We have changed this replacement and, as in *rmcr*, used the median value of brightness of the local region near the signature for such replacement. *Imgclean* (as *rmcr*) was also corrected not to consider pixels located close to the edges of the frame as CR events because these pixels does not contain an image itself. The sizes of these edges are presented in Table IV in (Hampton et al., 2005), but usually we considered the edges wider by 1-3 pixels (depending on the mode of the frame) than the edges from the table. For example, for mode 1 (1024*1024 pixels) Hampton et al. (2005) considered the edge of 8 pixels, but our analysis of images showed that sometimes even 11[th] line can contain some bright pixels which are not CR signatures.

On ITS and IR images there are a lot of pixels with high brightness which appeared at the same places on different images. The information about such 'warm' pixels can be extracted from a .fit file with a calibrated image. The 'warm' pixels were found based on analysis of many images. For removal of signatures of CRs on ITS and IR calibrated images (see sections 5-6), we modified *imgclean* in such a way that if the brightness of a 'warm' pixel is greater than the median value of close pixels, then it is replaced by this median value (in our studies we also used the code *rmwarm* that only makes the above replacement without searching for CR signatures). The search of CR signatures was made only for such modified ITS and IR images.

In our studies discussed in sections 6-7, we used also our code *checkcr* that finds CR signatures based on two dark images and on bad pixel maps (these maps were obtained by the DI team). A cluster of pixels is considered to be a signature of a CR if it presents only on one image and is located far from 'bad' pixels. We considered that this cluster must be separated by at least one pixel from any cluster on another image and from bad pixels. The distance criteria (similar to *ddilim*) is an input parameter and can be changed. Before using *checkcr*, we applied our code similar to *rmcr* to the .fit files that contained information about CR signatures, detected by a code another than *rmcr* (e.g., by *imgclean*), and to the file with bad pixels in order to find groups of pixels which are separated by a distance less than that corresponding to *ddilim*. Note that a cluster can consist of a different number of pixels on two images, but it can be considered as the same cluster if it satisfies the distance criterion.

## 3. Statistics of glitches of cosmic ray signatures on images

CR events are most easily detected on dark images. Our studies of statistics for CR events were based on analysis of raw dark images with the use of a code similar to *rmcr* for the case when all detected clusters are considered as CR signatures. Only pixels with digital numbers DN>370 for MRI and DN>390 for HRI were considered candidates for being CR events (i.e., signals more than ~15 DN above the bias level). The obtained statistics for CR signatures on visual and IR images was considered by Klaasen et al. (2006). In this section we discuss only the results important for choosing the input parameters of *rmcr* for studies non-dark images and for analysis of these images.

For most HRI and MRI visual images made during low solar activity at exposure time $t>4$ s, the number of CR events per second per square centimeter of CCD was about 2-4 (typically ~3), and generally there were no events consisting of more than $2t$ pixels. The above numbers can be greater at $t<4$ s.

For $t<0.2$ s the number of detected CR signatures can vary by a factor of several in different images with the same exposure time. It is caused by that the CR integration period is longer than the total image exposure time. Even for exposure time of about a few milliseconds, the number of CR events per square centimeter (or per quadrant consisted of 512 by 512 pixels) usually exceeded 2.

Most CR events in an image consist of not more than 4 pixels. The largest CR signatures have a linear form in contrast to the more circular form for star images. The brightest pixels belong to those CR events that consist of $N_p$~5-11 pixels. Small ($N_p \leq 3$) events usually consist of faint pixels, and large rays ($N_p>20$) are not bright.

At high solar activity, the CR event rate can increase by a factor of 5 compared to that at low activity, and long signatures of CRs can exceed $8.5t$. At $t=30$ s the maximum number of pixels in one long CR signature exceeded 200, while no CR events consisted of more than 45 pixels at $t=30$ s for images outside the period of solar flares. The ratio of the number of CR events consisting of $n$ pixels obtained at high solar activity to that at low solar activity was greater for greater $n$. For example, this ratio was greater at high solar activity than that in out-of-peak activity by a factor of about 1.5, 2, 3, 3.5, 7 for rays consisted of 1, 2, 3, 4, and 5 pixels, respectively. This suggests that events caused by energetic particles from the Sun tend to produce larger signatures than do interstellar CRs.

Based on comparison of CR signatures on dark and sky images, we can make two main conclusions that were used for construction of *rmcr*:

(1) Most CR events consist of a small number of pixels, while well-exposed star images are typically larger (especially for the out-of-focus HRI). At $t \geq 4$ s, almost all (>80%) 1-4 pixel charge clusters in typical sky images are CR signatures (excluding images of dense conglomerations of stars).

(2) Large CR events have a linear form in contrast to the more circular form of star images. We calculated the ratio $k_p = n_{pix}/(dxx^2+dyy^2)$ for different clusters, where *dxx* and *dyy* are the maximum differences of coordinates *x* and *y* in a charge cluster each increased by 1, and $n_{pix}$ is the total number of pixels in the cluster. At $n_{pix}>30$ we found that $k_p<0.17$ for all CR signatures on dark images and $k_p>0.17$ for all stars. However, when charge clusters consist of ≤10 pixels, it is difficult to distinguish between signatures of CRs and stars based on $k_p$.

**4. Removal of cosmic ray signatures on visual MRI and HRI images**

The effectiveness in recognizing CR signatures with the different codes varied for different kinds of images. Codes *imgclean, crfind,* and *di_crrej* often do not work normally with raw images. In this case, sometimes they had errors during their processing and didn't run to completion; sometimes they deleted a lot of arbitrary pixels of background. *Rmcr* does not have problems with raw images. Below we briefly discuss the performance of the above four codes, giving particular attention to those images for which the codes do not work well. A few examples of images with clusters detected as CR signatures by different codes were presented by Klaasen et al. (2006).

First we discuss the removal of CR signatures from MRI and HRI calibrated images. Analysis of CR signatures detected by different codes showed that all four considered codes detected too many pixels near the edge of a comet as CR signatures. Codes *imgclean, crfind,* and *di_crrej* may have problems with detection of all pixels of long CR events. They often delete too many pixels near or inside an image of a large bright star. *Imgclean* may recognize long CR signatures events worse than other codes, and it may delete many of small faint stars in conglomeration of stars (when the number of a few pixel objects corresponding to stars are greater than that for CR signatures). Nevertheless, *imgclean* was chosen by us for automatical removal of CR signatures. The main advantage of this code is that it always comes to the end and need not much time for calculations. *Crfind* and *di_crrej* sometimes were not able to make calculations to the end. *Rmcr* may need to change input default parameters for better work with a concrete image, and it works slow for the images when a large number of pixels are brighter than the considered level of brightness. *Imgclean* is a more reliable code if one needs to remove CR signatures automatically from a large number of images, but depending on a specific image and a specific problem, other codes can work better (e.g., sometimes *crfind* is the best when there is a large image of a comet).

For the calibrated images considered (dark images and images of conglomerations of stars) with maximum radiance of ~0.0001 W-m$^{-2}$-μm$^{-1}$-sr$^{-1}$, *di_crrej* and *crfind* did not work normally if we used the same default parameters for which these codes worked normally with calibrated images with maximum radiance of ~1 W-m$^{-2}$-μm$^{-1}$-sr$^{-1}$. For the small radiance case, they deleted a lot of pixels of background. We have not found parameter settings that work well at small radiances for these codes.

We suggest not to use any code for removal of small CR signatures on the frames made just after the impact if the expected number of CR signatures is not more than e.g. 3 (for images consisted of not more than 256*256 pixels and made with a short exposure time, e.g. less than 1 second) and an image of a comet occupies a considerable part of the frame. In this case the number of false CR signatures detected by any code near the edge of a comet may be greater than the number of real CR signatures. Rare long CR signatures on such frames can be deleted by the interactive code discussed in section 7.

**5. Removal of cosmic ray signatures on visual ITS images**

CR detection has proved more difficult in ITS images than in either MRI or HRI images. The problem may be because the values of the background DNs in an ITS raw dark image vary more than those for MRI and HRI images (by factors of 2 or more) due to the higher operating temperature and increased dark current of the ITS CCD. The difference between the median DN values for different quadrants of ITS dark images can exceed 40.

For raw ITS images only *rmcr* works. If we consider pixels with DN greater by 20 than the median value of DN for a quadrant, then the number of clusters recognized by *rmcr* as CR signatures on a dark image is greater by a factor of 5-10 than that for MRI and HRI at the same exposure time. Most of these clusters (>80%, and sometimes >90%) consisted of 1 pixel. The difference between DNs of these pixels and the median value of DNs was about the same as that for CR events on MRI images, so it is not possible to detect 1-pixel clusters as CR signatures on ITS images. The number of clusters consisted of $n≥4$ pixels was similar to the number of CR events on MRI images, and their mean DNs often exceeded that of 1-pixel clusters, so at $n \geq 4$ most of clusters probably were CR signatures. At $2 \leq n \leq 3$ about a half of clusters can be CR signatures, but usually it is not possible to conclude which of them are CR signatures, basing on their DNs. The number of charge clusters classified as CR signatures on a dark ITS image per second per quadrant was about twice greater than that for MRI and HRI if we consider only pixels with DN greater 50 above the median value for a quadrant, and in this case more than 80% of clusters consisted only of one pixel. The fraction of clusters detected as CR signatures and common for a pair of images (presumably, therefore, not true CR signatures) is about 5% of the clusters detected as CR events on one image when a threshold of 50 DN above the background is used.

None of the codes considered worked well with calibrated dark ITS images using their default parameter settings. The number of charge clusters consisting of $\leq 4$ pixels deleted by *imgclean* and *rmcr* was greater by a factor of several than even the expected number of CR events at the peak of solar activity, so most of the deleted clusters were not real CR signatures. *Crfind* and *di_ccrej* designated even more pixels as CR signatures. These excess CR detections appear to be due to the inadequacy of simply using a quadrant-mean dark current subtraction technique for the ITS. The excess CR detections tend to occur at the same pixel locations in all frames. This problem can be corrected by implementing a pixel-by-pixel dark current subtraction technique for ITS, and this work is in progress. While increasing parameters, which are multiplied in *imgclean* by the standard deviation of brightness of pixels in some region, by a factor of several (from the default values), we decreased the number of clusters considered as CR signatures by a factor of several, but this number was still greater by an order of magnitude than the expected number of CR events, and the difference with the expected number was not only for 1-3 pixel clusters, but also for greater clusters. Using *rmcr*, we obtain the number of clusters considered as CR signatures to be about the expected number of CR events, only if we consider $klim \geq 30$, i.e., by a factor of 10 greater than the default value of *klim*, but in this case the fraction of 1-pixel clusters among all clusters was greater by a factor of 2 than for MRI images.

The number of false CR signatures can decrease considerably if before applying *imgclean* we replace the brightness of 'warm' pixels (those bad pixels which are brighter than local background at different frames) by the median value of brightness of close pixels. However, even in this case for many ITS images the number of false CR signatures exceeded the expected number of CR signatures by a factor of several or more. For frames with a small image of a comet, there could be large groups of clusters detected as CR signatures and located far from a comet. Removal of such false CR signatures probably does not spoil an image with a comet, but only makes the brightness of background smoother. For frames with a large image of a comet, most of false CR signatures are located near the edge of the comet and sometimes on the comet itself. Removal of such false CR signatures can spoil the image.

The number of clusters recognized as CR signatures by *imgclean* on a dark ITS image and separated by more than one pixel from bad pixels (this check is made by our code, not by *imgclean*) can be about the number of expected CR signatures (in this case the number of CR signatures not coincided with bad pixels is greater by a factor of 1.5 than the number of CR signatures separated from bad pixels by more than one pixel) and can be less by a factor of 50 than the total number of clusters recognized as CR signatures by *imgclean* on an ITS image without replacement of brightness of warm pixels.

If an image of a comet is relatively small, then most of false CR signatures are close to warm pixels. In this case it is possible to find most of false CR signatures if we check the distances of clusters detected as CR signatures from warm pixels. If an image of a comet and/or coma in ITS frame is large, then *imgclean* detects too many (up to thousands, depending on the size of an image of a comet) false CR signatures even if we consider only clusters located far from warm pixels. In this case, the false CR events are located mainly near the edge of the comet and in coma. *Imgclean* can be used for studies of separate ITS frames without large images of a comet/coma with a subsequent check of distances of detected CR signatures from warm pixels (this check is not yet made automatically). In the case of automatic removal of CR signatures on ITS images, many false CR signatures can be removed, but may be such removal will not spoil images (at least, when there is no large image of a comet or coma) making the brightness of background of a comet smoother. The decision whether to remove CR signatures will depend on the aim of the use of the images.

For ITS images 256*256 or smaller, the number of real CR signatures on an image is very small (only a few clusters), so removal of small CR signatures on the images with small exposure time and small sizes will not improve the images even if it done correctly. We can look only for rare long CR signatures on such images.

## 6. Removal of cosmic ray signatures on infrared images

IR images have a much higher level of background noise than visual images due to the large signal produced by the SIM bench emission. In addition, the IR detector has many bad pixels that produce either far more or far less signal than a typical pixel when exposed to a uniform illumination field.

The most reliable way of detecting CR events in IR images is to difference two IR frames taken in succession to find the changes, most of which will be due to CR signatures if they are above the noise floor. Most IR data sets consist of multiple frames, either from a spatial scan or from repeated exposures of the same scene, and one can therefore take advantage of this internal redundancy to isolate CR events. To estimate the number of CR signatures, *imgclean* was run on two successive calibrated IR dark frames (including bad-pixel and flat-field corrections) and the number of charge clusters that were different between the two runs was determined. *Imgclean* classified about 2-3 events per $cm^2$ per second as CR signatures, consistent with the rates seen in visible images. Most of these CR charge clusters consist of only 1-4 pixels consistent with the event rate seen in CCD images. For the above estimates, we did not consider clusters close to bad pixels, and did not consider clusters presented on both frames, as the number of false CR signatures detected by *imgclean* on one IR image is greater by a factor of several than the number of real CR signatures. For studies of CR signatures on infrared images, we used our code *checkcr,* which finds CR signatures based on comparison of two images and bad pixel maps. This code was applied to CR signatures found by *imgclean* on separate frames. Long CR signatures were seen less often on IR images than on visual images (at the same exposure times).

The number of charge clusters classified as CR signatures by *imgclean* run on a single calibrated IR dark image is typically greater by a factor of about 5 than the number of clusters on this image that are not present on an immediately sequential identical neighboring image. Most of the clusters deleted by *imgclean* from a calibrated IR dark image can also be found in the same locations on other images and are not true CR signatures. In many cases, they contain known bad pixels in the detector. Running a "bad-pixel" reclamation routine prior to *imgclean* eliminates about half of these false detections.

In our test runs with calibrated IR dark images, only *imgclean* worked without problems, but it detected too many false CR signatures even after bad-pixel and flat-field correction. *Crfind* and *di_crrej* had problems running to completion with all parameter settings we tried. *Rmcr* needs a relatively constant background, which is not necessarily the case for IR frames. These codes may be useful for removing long CR signatures if they are present on IR images, but in this case one must be careful not to remove any long lines that belong to the spectrum or to stars (i.e., remove only oblique CR signatures, but not vertical or horizontal lines).

None of the codes have yet been made to work successfully on non-dark (either calibrated or raw) IR images; the number of false CR signatures can exceed the real number by a factor of 100 in runs made to date. The number of false CR signatures can become smaller by a factor of several if we use an image where brightness of bad pixels is replaced by the median value of local brightness, but for some images, the number of false CR signatures can still exceed the real number of CR signatures by a factor of several tens. For other images, it can exceed the estimated number of CR signatures by less than a factor of 2. Work is ongoing to develop CR detection approaches that are generally applicable to IR images.

We do not recommend to remove CR signatures automatically from all IR images because the number of false CR signatures can exceed the number of real CR signatures by a factor of several or more. For IR and ITS images, it may be useful to replace the brightness of bad pixels by the median neighboring brightness and do it as a part of pipeline. The replacement can be done similar to that discussed in section 2, but it may be useful to replace the brightness of all bad pixels, not only of those which are brighter than the local median brightness. Both, too bright and less bright pixels can spoil an image. Also negative brightness of all pixels which had such brightness can be replaced by zero brightness for all images.

## 7. Interactive code for removal of cosmic ray signatures

We worked out the interactive code *imr* which allows one to control manually the removal of CR signatures on images (subversion *imrt* produces not only output array, as *imr*, but also a few .fit files). The first step of the interactive code is to replace the brightness of 'warm' pixels by the local median brightness (this step can be deleted for MRI and HRI images). For ITS images, 'warm' pixels can provide a great number of false detected CR signatures. The code can show the clusters which *imgclean* detects as CR events. It restores the edges of an image after *imgclean* (see section 2). Based on an exposure time and a size of the image, the code calculates the expected number of CR signatures.

The code *imr* allows a user to choose the rectangles "A" on a considered image where glitches recognized by *imgclean* as CR signatures are ignored. In other rectangles "B" which can be chosen by the user, the brightness of some pixels is replaced by the local median brightness if the brightness of these pixels is greater by some factor *kdif* (the default value of *kdif* is 2) than the median brightness.

The rectangles are marked by clicking a mouse at two apices. If one knows in advance the coordinates of corrected regions, he can print the coordinates. In the latter case it is possible to work with images without seen them (i.e., for remote access to an IDL computer).

The result of the work of the code is the following: (1) inside rectangles "A", an initial image (i.e., before *imgclean*) is remained; (2) inside rectangles "B", the brightness is changed for clusters detected by *imgclean* as CR signatures and for other pixels which brightness differed by a factor of *kdif* from the local median brightness; (3) outside rectangles "A" and "B", the brightness is changed for the clusters detected by *imgclean* as CR signatures.

*Imgclean* often detects false CR signatures near the edge of a comet. So, one can choose several rectangles "A" which include such false CR signatures. These rectangles can occupy the edges of the comet, or one can take one larger rectangle which includes a whole image of the comet and its edges. On some ITS images, there are regions where the brightness of pixels corresponding to background varied much for close pixels, and therefore *imgclean* can detect a great number of false CR signatures in these regions. Rectangles "A" can also include such regions. We recommend to use rectangles "B" in order to delete those glitches of CRs which were not entirely deleted by *imgclean* (often *imgclean* can not delete all pixels of a long CR event). Also one may want to delete stars if he wants. In the above 'wide' way in *imr*, one can use any number of rectangles and can make many corrections of the image. This version is recommended if one wants to edit an image carefully.

A 'quick' way in the code allows a user to choose a rectangle "C" inside which clusters recognized by *imgclean* as CR signatures are not considered as CR events (i.e., inside this rectangle, there are no changes compared with the original image). Outside this rectangle, the brightness of all pixels which differed from the median value of local brightness by a factor of more than *kdif* is replaced by the median value. Inside rectangle "C" we make the same calculations as for the rectangles "A", and outside the rectangle "C", we do the same as for rectangles "B". It is suggested to choose a rectangle "C" so that it will include an image of a comet and some region near the edges of the comet (where usually *imgclean* detects false CR signatures; for images with outbursts this region can be greater).

The 'quick' way is recommended when one needs to consider quickly a large number of images. For example, it can be used for small frames with a small image of a comet. Outside the rectangle "C", all stars are also deleted. Note that the 'quick' way can work (after rewriting the code) without using *imgclean*, but in this case outside rectangles "C" we will change the brightness only of the pixels which are differed from the local median brightness. For frames without a comet, there is no sense to use a 'quick' way in *imr*.

The interactive code allows a user to delete those long CR signatures which are not recognized by *imgclean* and to prevent consideration of the pixels close to the edge of a comet as CR events, i.e., to solve two main problems which prevent to obtain good final MRI and HRI images with *imgclean*. The interactive code was used by B. Carcich for corrections of DI images presented at the DI web-site.

*Imr* can be used for removal of long oblique CR signatures (vertical and horizontal segments are not caused by CRs) on ITS and IR images. At small exposure times, the probability of long CR signatures is small. It may be possible to check only ITS and IR images with exposure time of at least a few seconds and to find those images where long CR signatures are present. For such images one can delete only long CR signatures (first take a rectangle "A" to be equal to the whole image in order to not consider the CR signatures found by *imgclean*; then put rectangles "B" around long CR signatures in order to delete these signatures).

## 8. Conclusions

We analyzed the work of several codes (*imgclean, crfind, di_crrej,* and *rmcr*) written by several authors and recognizing cosmic ray (CR) signatures on one image. Images made by different cameras (HRI, MRI, and ITS) during the flight of Deep Impact to Comet Tempel 1 were considered. For automatic removal of CR signatures on many visual images, we recommend to use *imgclean* (with corrected replacement of the brightness of detected CR signatures), but for analysis of concrete images, other codes can give better results. *Imgclean* detects false CR signatures near the edge of a comet, and it often does not recognize all pixels of long CR signatures, but other codes considered sometimes does not come to the end. Our code *rmcr* is the only code among mentioned above which allows to work with raw images. In some cases (e.g., for removal of CR signatures near a bright star), it works better than other above codes, but for many calibrated images it has no advantages.

For most HRI and MRI visual images made during low solar activity at exposure time $t>4$ s, the number $N_{sc}$ of clusters of bright pixels on an image per second per $cm^2$ of CCD was about 2-4, both for dark and normal sky images. At high solar activity, $N_{sc}$ sometimes exceeded 10. The ratio of the number of CR signatures consisting of *n* pixels obtained at high solar activity to that at low solar activity was greater for greater *n*.

Due to higher variations of brightness of background, at default parameter settings, all the codes considered detected too much false CR signatures on ITS images. Clusters consisted of less that 4 pixels, usually can not be surely identified as CR signatures on ITS CCDs at any parameters, as the brightness for such small CR signatures is low enough.

The number of clusters detected as CR signatures on a single infrared image is by at least a factor of several greater than the actual number of CR signatures, but the number of clusters based on analysis of two successive dark frames is in agreement with an expected number of CR signatures. Some false CR signatures include bright pixels presented on different infrared images.

Our interactive code *imr* allows a user to choose the regions on a considered image where glitches detected by *imgclean* as CR signatures are ignored. In other regions chosen by the user, the brightness of some pixels is replaced by the local median brightness if the brightness of these pixels is greater by some factor than the median brightness.

**Acknowledgement**s

The authors are thankful to B. Carcich, M. Desnoyer, and D. Linder for useful advices and discussions. The work was made possible to support from NASA for the Deep Impact project.

**References**

Hampton, D.L., Baer, J.W., Huisen, M.A., Varner, C.C., Delamere, A., Wellnitz, D.D., A'Hearn, M.F., Klaasen, K.P. Deep Impact: An overview of the instrument suite for the Deep Impact mission, Space Science Reviews 117, 43-93, 2005.
A'Hearn, M.F., Belton, M.J.S., Delamere, W.A., Kissel, J., Klaasen, K.P., McFadden, L.A., Meech, K.J., Melosh, H.J., Schultz, P.H., Sunshine, J.M., Thomas, P.C., Veverka, J., Yeomans, D.K., Baca, M.W., Busko, I., Crockett, C.J., Collins, S.M., Desnoyer, M., Eberhardy, C.A., Ernst, C.M., Farnham, T.L., Feaga, L., Groussin, O., Hampton, D., Ipatov, S.I., Li, J.-Y., Lindler, D., Lisse, C.M., Mastrodemos, N., Owen, W.M., Jr., Richardson, J.E., Wellnitz, D.D., White, R.L. Deep Impact: Excavating Comet Tempel 1. Science 310, 258-264, 2005.
Klaasen, K.P., A'Hearn, M.F., Baca, M.W., Delamere, W.A., Desnoyer, M., Farnham, T.L., Groussin, O., Hampton, D., Ipatov, S.I., Li, J.-Y., Lisse, C.M., Mastrodemo, N., McLaughlin, S., Sunshine, J.M., Thomas, P.C., Wellnitz, D.D. Deep Impact Instrument Calibration. Optical Engineering, submitted, 2006.